\documentclass[twocolumn,showpacs,preprintnumbers,superscriptaddress]{revtex4}
\usepackage{amsmath}
\usepackage{dcolumn}
\usepackage{bm}
\usepackage{graphicx}

\begin{document}

\title{Entanglement of separate nitrogen-vacancy centers coupled to a
whispering-gallery mode cavity}
\author{W. L. Yang}
\affiliation{State Key Laboratory of Magnetic Resonance and Atomic
and Molecular Physics, Wuhan Institute of Physics and Mathematics,
Chinese Academy of Sciences, Wuhan 430071, China }
\author{Z. Y. Xu}
\affiliation{State Key Laboratory of Magnetic Resonance and Atomic
and Molecular Physics, Wuhan Institute of Physics and Mathematics,
Chinese Academy of Sciences, Wuhan 430071, China }
\affiliation{Graduate School of the Chinese Academy of Sciences,
Beijing 100049, China}
\author{M. Feng}
\altaffiliation{mangfeng@wipm.ac.cn} \affiliation{State Key
Laboratory of Magnetic Resonance and Atomic and Molecular Physics,
Wuhan Institute of Physics and Mathematics, Chinese Academy of
Sciences, Wuhan 430071, China }
\author{J. F. Du}
\altaffiliation{djf@ustc.edu.cn} \affiliation{Hefei National
Laboratory for Physics Sciences at Microscale and Department of
Modern Physics, University of Science and Technology of China,
Hefei, 230026, China}

\begin{abstract}
We present a quantum electrodynamical model involving nitrogen-vacancy
centers coupled to a whispering-gallery mode cavity. Two schemes are
considered to create W state and Bell state, respectively. One of the
schemes makes use of Raman transition with the cavity field virtually
excited and the other enables Bell state preparation and quantum information
transfer by virtue of dark state evolution and adiabatic passage, which is
tolerant to ambient noise and experimental parameter fluctuations. We
justify our schemes by considering the experimental feasibility and
challenge using currently available technology.
\end{abstract}

\maketitle

\section{Introduction}

The diamond nitrogen-vacancy (NV) center consisting of a substitutional
nitrogen atom and an adjacent vacancy has attracted considerable attention
since the first report of optically detected magnetic resonance on single NV
center in 1997 \cite{first}. Due to sufficiently long electronic spin
lifetime as well as the possibility of coherent manipulation at room
temperature \cite{Childress}, the NV center is considered as a promising
building block for room-temperature quantum computing in the future \cite%
{To,room-tem, shi}.

Since the electronic spins could be well initialized and manipulated in
optical fashion, the qubit readout and gating regarding single-spin state
have been achieved in individual NV centers \cite{jele1}. By virtue of the
hyperfine couplings with the paramagnetic nuclei in the vicinity of the
electron spin, i.e., $^{13}$C \cite{register}, $^{14}$N \cite{Han}, and $%
^{15}$N \cite{Jac}, currently available techniques have demonstrated the
quantum information storage and retrieval between electronic and the nuclear
spins \cite{register}. This technique also enables rapid and high-fidelity
readout of quantum information from the electron spin \cite{readout}.
However, coherence between electron and nuclear spin qubits is restricted to
the case of a few qubits owing to the limited number of nuclear spins
individually addressable in frequency space \cite{Neu1,Neu2}. So for
scalability, it is necessary to develop methods of coupling distant NV
centers.

We have noticed a recent experiment to entangle a pair of separate NV
centers within a diamond based on magnetic dipolar coupling \cite{Nat}. But
this idea is pretty hard for scalability. For distant NV centers with
magnetic dipolar coupling unavailable, the best way for spin-spin
entanglement seems to make use of parity projection by detecting the emitted
photons relevant to different spin states \cite{Review}. However, the NV
centers, although similar to the atomic cases, only allow linearly polarized
radiations in the laser excitation, which makes coincident detection of
emitted photons unavailable. In addition, to effectively produce
entanglement of NV centers by parity projection of photons, we require that
the `which-path' information be removed due to interference after the
photons go through the beam splitter. But experimental reports so far have
shown that $96\%$ of the emitted photons reside in broad photon sidebands to
the resonant zero phonon line (ZPL) at 637 nm even in cryogenic situation
\cite{Review}. This implies that the most photons emitted from the NV
centers could not effectively interfere in the beam splitter.

Alternatively, the entanglement of separate NV centers could be achieved by
coupling to the same cavity mode. In a recent publication \cite{APL}, we
proposed an idea to entangle more than two NV centers by one step of
implementation, based on coupling to a microsphere cavity. The key idea of
that proposal is the employment of the spin singlet state $^{1}A$ to encode
a qubit. To our knowledge, however, this metastable $^{1}A$ state, although
investigated from $C_{3v}$ group theory considerations \cite{Aco} and other
aspects, has not yet been fully understood \cite{Rog}. As a result, most of
the present work for quantum information processing with NV centers encode
qubits only in the sublevels of the ground state $^{3}A$.

We focus in this work on entangling distant NV centers without employing the
state $^{1}A$. Specifically, we encode qubits in two of the ground state
splittings, and the excited states are auxiliary with spontaneous emission
effectively suppressed during our operation. The key point of our idea is to
present a generalized Jaynes-Cummings model involving a quantized
whispering-gallery mode (WGM) and $N$ identical NV centers. WGM
microcavities are of typically dielectric rotational-symmetry structures
with WGMs traveling around the curved boundary and confined by continuous
total internal reflection \cite{Val}. In particular, the technological
advance has made it available to have strong light-matter coherent coupling
in WGM resonators with smaller mode volume $V_{m}$ and extremely
high-quality factor $Q$ \cite{Buck,WG}. Recent experimental progresses about
the nanocrystal-microsphere system also provide experimental evidence for
strong coupling between NV centers and the WGM of silica microsphere \cite%
{park}, polystyrene microsphere \cite{sch}, and gallium-phosphide microdisk
\cite{bar}, respectively. In addition, a latest experiment has demonstrated
the technique for deterministically coupling a single NV center to a
photonic crystal cavity \cite{Englund}. So far there have been much
development in WGM cavities with the forms, such as the microtoroidal \cite%
{tor}, microcylinders \cite{lin}, microdisks \cite{disk}, and microspheres
\cite{W1}.

We will show how to generate $W$ state \cite{W} and Bell states for these
distant NV centers in such a composite nanocrystal-microsphere system. The
main results of this work are twofold: First, by virtue of Raman
transitions, we show the possibility with virtual-photon-induced excitation
in a large detuning case to generate multipartite W state with separate NV
centers in different diamond nanocrystals, where the growth of the qubit
number corresponds to the decrease of the operational time. Secondly,
resorting to adiabatic passage technology \cite{ada,CP}, we create Bell
states of any pairs of qubits via dark-state evolution, which is robust to
the cavity decay.

\begin{figure}[tbph]
\centering
\includegraphics[width=8cm]{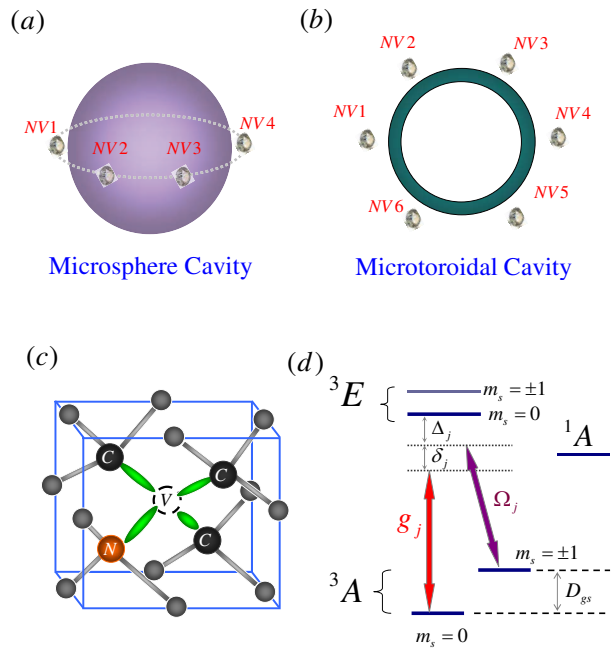} \caption{(a)
Schematic setup in the microsphere cavity case, where $N$ identical
NV centers in diamond nanocrystals are equidistantly attached around
the equator of a single fused-silica microsphere cavity. (b)
Schematic setup in the microtoroidal cavity case, where $N$
identical NV centers in the vicinity of the microtoroidal cavity are
able to interact with the WGM via the evanescent field. (c)
Structure of the NV color center in the diamond lattice, consisting
of a substitutional nitrogen (N) and a
neighboring vacancy (V). (d) Level diagram for the $j$-th NV center, where $%
\Delta _{j}$ and $\protect\delta _{j}$ are detunings, $g_{j}$ is the
coupling strength between NV center and WGM, and $\Omega _{j}$ is the
coupling strength between NV center and the laser pulse. $D_{gs}=$2.88 GHz
is the zero-field splitting between the ground state sublevels $m_{s}=0$ and
$m_{s}=\pm 1$ ($m_{s}=\pm 1$ are degenerate at zero magnetic field due to $%
C_{3v}$ symmetry) of the NV center. We encode qubits in the subspace spanned
by the down state $m_{s}=0$ and the up state $m_{s}=-1.$}
\end{figure}

\section{Entanglement generation by Raman configuration}

\subsection{Effective Hamiltonian}

As sketched in Fig. 1(a) and 1(b), $N$ identical NV centers, respectively,
in $N$ separate diamond nanocrystals could be strongly coupled to the WGM of
a microsphere cavity or of a microtoroidal cavity. The NV center is a point
defect in the diamond lattice, which consists of a nearest-neighbor pair of
a nitrogen-atom impurity substituting a carbon atom and a adjacent carbon
vacancy, as shown in Fig. 1(c). In the case of microspheres, WGM can be
characterized by angular $l$, azimuthal $m$, and radial $s$ numbers. High
values of $Q$ usually correspond to the modes with $l\gg 1$. Of greatest
interest is the so-called fundamental WGM ($s=1, l=m$), whose field is
concentrated in the vicinity of the equatorial plane of the sphere. It is
believed that these modes can be selectively excited by coupling to a
tapered fiber \cite{Spi2}. Like in \cite{APL}, our proposal is also based on
recent experimental and theoretical progresses, i.e., the possibility of $%
\Lambda $-type configuration of the optical transition in NV center system
\cite{lambda} and the considerable enhancement of the ZPL by embedding the
NV centers in some cavities \cite{cavity1}. But differently, the
entanglement of the NV centers is achieved without involving the metastable $%
^{1}A$ state. In our case, we assume that each NV center located in a
diamond nanocrystal is attached around the equator of a single fused-silica
microsphere cavity \cite{W1} or microtoroidal cavity \cite{tor}. This
composite nanocrystal-microsphere system takes advantage of the exceptional
spin properties of NV centers as well as the ultrahigh quality factor $Q$ $%
(\geq 10^{8}$ even up to 10$^{10})$, very small volume ($V_{m}\leq 100$ $\mu
m^{3}$) and simple fabrication technique of the cavity \cite{Ar}. For
convenience of description, we will below mention the WGM, but not relating
the mode to any concrete cavity.

By combining laser pulses with carefully timed interaction with the WGM, one
can model the NV center as a $\Lambda $-type three-level system as shown in
Fig. 1(d), where the states $\left\vert ^{3}A,m_{s}=0\right\rangle $ and $%
\left\vert ^{3}A,m_{s}=-1\right\rangle $ serve as the logical states $%
\left\vert 0\right\rangle $ and $\left\vert 1\right\rangle $ of the qubit,
respectively, and the state $\left\vert ^{3}E,m_{s}=0\right\rangle $ is
labeled by the state $\left\vert e\right\rangle $. In our case, the WGM with
frequency $\omega _{c}$ is far-off resonant from the transition $\left\vert
0\right\rangle \iff \left\vert e\right\rangle $ (with the frequency $\omega
_{e0}$), and the levels $\left\vert 1\right\rangle $ and $\left\vert
e\right\rangle $ (with transition frequency $\omega _{e1}$) are coupled by a
largely detuned laser with frequency $\omega _{L}$ and polarization $\sigma
^{+}$ \cite{opt,lambda}. The NV centers are fixed and apart with the
distance much larger than the wavelength of the WGM, interacting
individually with laser beams. So the direct coupling between NV centers is
negligible. Assuming that the detuning $\Delta_{j}$ is sufficiently larger
than the coupling strength $g_{j}$ and $\Omega_{j},$ the excited state $%
\left\vert e\right\rangle $ can be adiabatically eliminated. Thus quantum
logic gates and multipartite entangled states are available in the subspace
spanned by $\left\vert 0\right\rangle$ and $\left\vert 1\right\rangle$.
Using the rotating-wave approximation (RWA), the Hamiltonian in the
interaction picture can be written in units of $\hbar =1$ as \cite{FM}
\begin{equation}
H_{I}=\sum\nolimits_{j=1}^{N}\eta _{j}[a^{+}\sigma _{j}^{-}e^{-i\delta
_{j}t}+a\sigma _{j}^{+}e^{i\delta _{j}t}],
\end{equation}%
where $\sigma _{j}^{+}=\left\vert 1_{j}\right\rangle \left\langle
0_{j}\right\vert $, $\sigma _{j}^{-}=\left\vert 0_{j}\right\rangle
\left\langle 1_{j}\right\vert $, and $a^{+}(a)$ is the creation
(annihilation) operator of the WGM field. $\eta _{j}=g_{j}\Omega _{j}(\frac{1%
}{\Delta _{j}+\delta _{j}}+\frac{1}{\Delta _{j}})$ with $\Delta _{j}=\omega
_{e1,j}-\omega _{L,j}$ and $\delta _{j}=\omega _{e0,j}+\omega_{L,j}-\omega
_{e1,j}.$ For simplicity, we assume that the detuning $\delta _{j}$ and the
interaction term $\eta _{j}$ are identical for each qubit, that is, $\eta
_{j}=\eta ,$ and $\delta _{j}=\delta .$ In the case of $\delta \gg \eta ,$
there is no energy exchange between the WGM and NV centers. If the quantized
WGM is initially in the vacuum state, the effective Hamiltonian could be
simplified as \cite{Wu}
\begin{equation}
H_{eff}=\gamma \lbrack \sum\nolimits_{j=1}^{N}\left\vert 1_{j}\right\rangle
\left\langle 1_{j}\right\vert +\sum\nolimits_{j,k=1,j\neq k}^{N}\sigma
_{j}^{+}\sigma _{k}^{-}],
\end{equation}
where the first term corresponds to the dynamical energy shift regarding the
level $\left\vert 1\right\rangle $, and the photon-dependent energy shift of
the level $\left\vert 0\right\rangle$ is removed due to the vacuum state of
the cavity. The rest terms in Eq. (2) denote the coupling between any pair
of NV centers through the WGM, and $\gamma =\left\vert \eta ^{2}/\delta
\right\vert $ is the effective coupling strength for the energy conversing
transition $\left\vert 1_{1}0_{2}\right\rangle \iff \left\vert
0_{1}1_{2}\right\rangle$.

\begin{figure}[tbph]
\centering
\includegraphics[width=6cm]{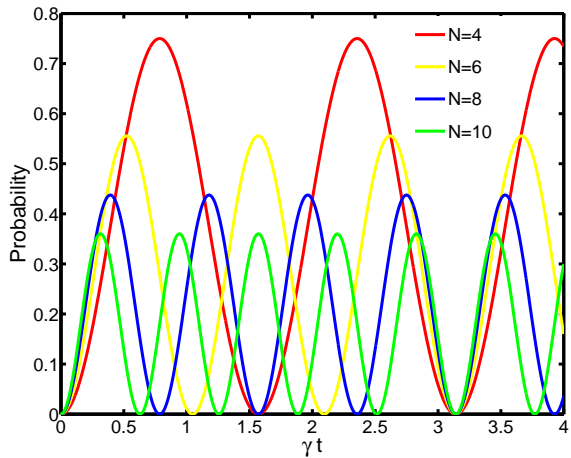} \caption{The
probability of obtaining the state $\left\vert W\right\rangle
_{N-1}$ versus $\protect\gamma t.$}
\end{figure}

\subsection{Creation of W state}

Let us first consider the creation of $N$-qubit $W$ state for NV centers. If
the first ($N-1$) NV centers are initially prepared in the state $\left\vert
00\cdot \cdot \cdot 0\right\rangle_{1,2, \cdot \cdot \cdot, N-1}$ and only
the $N$-th NV center is initially prepared in state $\left\vert
1\right\rangle_{N}$, one can easily get the following time-dependent state
evolution,
\begin{equation}
\left\vert \Psi (t)\right\rangle =C_{1}\left\vert 00\cdot \cdot \cdot
0\right\rangle _{1,2,\cdot \cdot \cdot ,N-1}\left\vert 1\right\rangle
_{N}+C_{2}\left\vert W\right\rangle _{N-1}\left\vert 0\right\rangle _{N},
\end{equation}
with the coefficients $C_{1}=(e^{-iN\gamma t}+N-1)/N$ and $C_{2}=\sqrt{N-1}%
(e^{-iN\gamma t}-1)/N$. $\left\vert W\right\rangle _{N-1}=(1/\sqrt{N-1})$ $%
\left\vert N-2,1\right\rangle $ is the generalized form of the $W$ state,
which denotes the symmetric state involving $(N-2)$ zeroes and $1$ one.
According to Eq. (3), the state of other $(N-1)$ NV centers will surely
collapse into the state $\left\vert W\right\rangle _{N-1}$ in the case of
the measurement on the $N$-th NV center being $\left\vert
0\right\rangle_{N}. $ As a result, we can find that the probability of
obtaining the state $\left\vert W\right\rangle _{N-1}$ is $\left\vert
C_{2}\right\vert ^{2}/(\left\vert C_{1}\right\vert ^{2}+\left\vert
C_{2}\right\vert ^{2})$, which gets to maximum $P_{\max
}=4(N-1)/[4(N-1)+(N-2)^{2}]$ at $t_{N}=(2k+1)\pi /N\gamma $ with $k$
non-negative integers.

Fig. 2 shows that the gating time $t_{N}$ is inversely proportional to the
qubit number $N$, so is the maximal probability $P_{\max }.$ In this
composite nanocrystal-microsphere system, the coupling between the NV center
and WGM could reach $g_{\max }=\Gamma _{0}\left\vert \vec{E}(r)/\vec{E}%
_{\max }\right\vert \sqrt{V_{a}/V_{m}}$ \cite{WG}, where $\left\vert \vec{E}%
(r)/\vec{E}_{\max }\right\vert $ is the normalized electric field strength
at the location $r$, and $V_{a}=3c\lambda ^{2}/4\pi \Gamma _{0}$ denotes a
characteristic interaction volume with $\lambda $ the transition wavelength
between the states $\left\vert e\right\rangle $ and $\left\vert
0\right\rangle $, $\Gamma _{0}$ the spontaneous decay rate of the excited
state $\left\vert e\right\rangle $ and $c$ the speed of light. Using the
values $\lambda =$637 nm, $\Gamma _{0}=$2$\pi \times 83$ MHz \cite{NJ}, $%
V_{m}=$100 $\mu m^{3}$, and $\left\vert \vec{E}(r)/\vec{E}_{\max
}\right\vert =$1/6, we have the maximal coupling $g_{\max }\simeq $2 $\pi
\times 1$ GHz, and the other experimental parameters can be adjusted as $%
\Omega_{j}=2\pi \times $ 100 MHz, $\Delta _{j}=2\pi \times $ 10 GHz, and $%
\delta _{j}=2\pi \times $ 100 MHz. Provided $\eta _{j}=2\pi \times 20$ MHz,
we have $\gamma =\left\vert \eta ^{2}/\delta \right\vert =2\pi \times 4$
MHz, and the operation time $t_{N}$ to be 0.0313 $\mu$s, 0.0208 $\mu$s, and
0.0156 $\mu$s in the case of $N=4$, $N=6$, and $N=8$, respectively.

\begin{figure}[tbph]
\centering
\includegraphics[width=8cm]{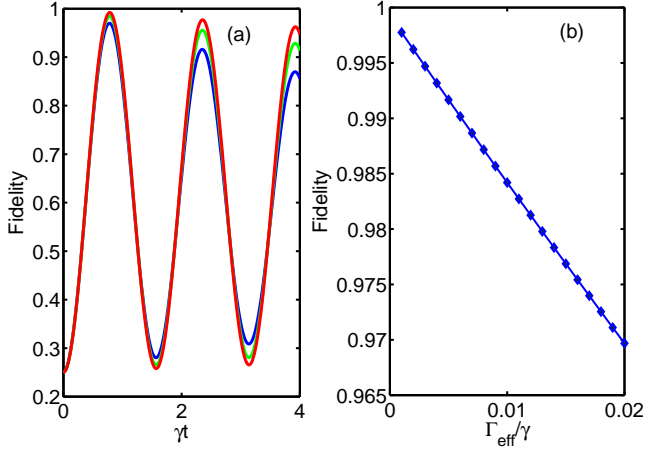} \caption{(a) The
fidelity of the state $\left\vert W\right\rangle _{4}$ versus
$\protect\gamma t$, where the blue, green and red curves correspond
to $\Gamma _{eff}=\protect\gamma /50,$ $\protect\gamma /100,$ and $\protect%
\gamma /200,$ respectively. (b) The fidelity of the state $\left\vert
W\right\rangle _{4}$ versus $\Gamma _{eff}/\protect\gamma $, where the
gating time is $\protect\pi /4\protect\gamma $.}
\end{figure}

\subsection{Estimate of decoherence}

We now consider the influence due to decoherence, which results from the
effective spontaneous emission from the states $\left\vert 1\right\rangle $
to $\left\vert 0\right\rangle $. Here we have neglected the WGM decay
because the cavity decay rate could be $\kappa =\omega _{e0}/Q=2\pi \times
0.47$ MHz in the case of $Q=10^{9},$ which is much smaller than the
effective coupling rate $\gamma $. The characteristic spontaneous emission
rate $\Gamma _{eff}$ regarding the states $\left\vert 1\right\rangle$ and $%
\left\vert 0\right\rangle$ could be estimated as $\Gamma_{0}\Omega
_{j}g_{j}/\Delta_{j}^{2}$ \cite{Spo}, where $\Gamma_{0}$ is the spontaneous
decay rate of the excited state $\left\vert e\right\rangle $ \cite{exp1}. So
the evolution of the system is described by the Lindblad equation \cite{Ma}
\begin{equation}
\dot{\rho}=-i[H,\rho ]+\Gamma _{eff}(2\sigma ^{-}\rho \sigma ^{+}-\sigma
^{+}\sigma ^{-}\rho -\rho \sigma ^{+}\sigma ^{-}).
\end{equation}%
Fig. 3 shows the fidelity of the state $\left\vert W\right\rangle _{4}$ when
the spontaneous decay is considered. With increase of $\Gamma _{eff}$, the
fidelity decreases accordingly. However, our scheme can still achieve a high
fidelity as long as the spontaneous decay is weak. In a realistic
experiment, the situation would be more complicated than our consideration
above. As a result, to carry out our scheme with high efficiency and high
fidelity, we have to suppress above mentioned imperfect factors as much as
we can.

\section{Entanglement generation by adiabatic passage of dark states}

\subsection{Dark State Evolution}

Under Raman resonance conditions between two Zeeman sublevels of the ground
state, we focus on generating Bell state with any pairs of NV centers (e.g.,
$A$ and $B$) in the nanocrystal-microsphere system, via adiabatic passage of
the dark states \cite{Parkins}. The adiabatic passage \cite{add} is a useful
and robust technique for quantum-state manipulation. It is a method of using
two time-separated but partially overlapping pulses in the counterintuitive
sequence to produce complete population transfer or an arbitrary coherent
superposition between the initial and final states.

The difference from the above section is that the WGM and the laser pulses
should be in resonance with the transitions $\left\vert 0\right\rangle \iff
\left\vert e\right\rangle $ and $\left\vert 1\right\rangle\iff\left\vert
e\right\rangle$, respectively, as shown in Fig. 4.

\begin{figure}[tbph]
\centering
\includegraphics[width=8cm]{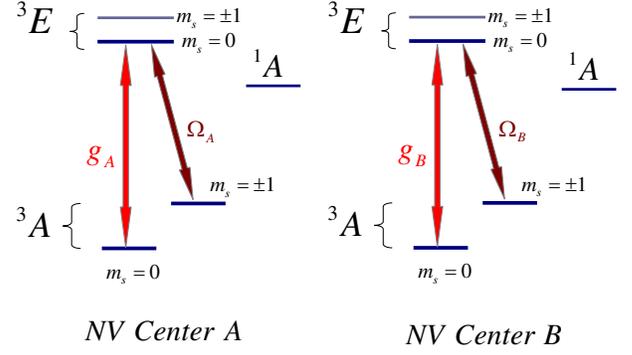} \caption{Level
diagram for two NV centers $A$ and $B$, where $g_{j}$ ($j=A$ and
$B$) is the coupling constant due to the WGM, and $\Omega _{j}(t)$
is the Rabi frequencies relevant to laser pulses.}
\end{figure}

The interacting Hamiltonian, after RWA, has the form
\begin{equation}
H_{I}^{^{\prime }}=\sum\nolimits_{j=A,B}[g_{j}a^{+}\left\vert
0_{j}\right\rangle \left\langle e_{j}\right\vert +\Omega _{j}(t)\left\vert
e_{j}\right\rangle \left\langle 1_{j}\right\vert +H.c.],
\end{equation}%
by which the Bell state is generated following the procedure below: (i) The
NV centers $A$ and $B$ are prepared in the initial state $\left\vert
1_{A}\right\rangle \left\vert 0_{B}\right\rangle $, and the WGM is initially
prepared in vacuum state $\left\vert 0_{c}\right\rangle $; (ii) We set
initially the Rabi frequencies $\Omega _{A}\ll \Omega _{B}$, and then we
adiabatically decrease $\Omega _{B}$ while increasing $\Omega _{A}$, driving
the system into a superposition of Zeeman sublevels which by quantum
interference are decoupled from the excited state and form a dark state $%
\left\vert D\right\rangle$ (defined later); (iii) We go on adiabatically
decreasing $\Omega_{B}$ while increasing $\Omega_{A}$ until $\Omega _{A}=$ $%
\Omega_{B}=\Omega_{0}$, we reach the state $\left\vert D_{f}\right\rangle $
and then turn off the laser pulses. These steps can be briefly described as
\begin{equation}
|1_{A}\rangle |0_{B}\rangle |0_{c}\rangle \underrightarrow{\text{(ii)}}%
||D\rangle \underrightarrow{\text{(iii)}}|D_{f}\rangle ,
\end{equation}%
where
\begin{eqnarray}
&&\left\vert D\right\rangle =\tilde{N}^{^{\prime }}[\Omega
_{B}(t)g_{A}\left\vert 1_{A}\right\rangle \left\vert 0_{B}\right\rangle
\left\vert 0_{c}\right\rangle +\Omega _{A}(t)g_{B}\left\vert
0_{A}\right\rangle \left\vert 1_{B}\right\rangle \left\vert
0_{c}\right\rangle  \notag \\
&&-\Omega _{A}(t)\Omega _{B}(t)\left\vert 0_{A}\right\rangle \left\vert
0_{B}\right\rangle \left\vert 1_{c}\right\rangle ]
\end{eqnarray}
is a dark state regarding the Hamiltonian in Eq. (5), and
\begin{eqnarray}
&&\left\vert D_{f}\right\rangle =\tilde{N}[\Omega _{0}(t)(g_{A}\left\vert
1_{A}\right\rangle \left\vert 0_{B}\right\rangle \left\vert
0_{c}\right\rangle +g_{B}\left\vert 0_{A}\right\rangle \left\vert
1_{B}\right\rangle \left\vert 0_{c}\right\rangle )  \notag \\
&&-\Omega _{0}^{2}(t)\left\vert 0_{A}\right\rangle \left\vert
0_{B}\right\rangle \left\vert 1_{c}\right\rangle ]
\end{eqnarray}
is the final state with $\tilde{N}$ and $\tilde{N}^{^{\prime }}$ being
normalization factors. In the case of $g_{A}=g_{B}=g_{0}$, the final state $%
\left\vert D_{f}\right\rangle$ is simplified to
\begin{equation}
\left\vert D_{f}\right\rangle =\tilde{N}[\sqrt{2}\Omega
_{0}(t)g_{0}\left\vert Bell\right\rangle _{AB}\left\vert 0_{c}\right\rangle
-\Omega _{0}^{2}(t)\left\vert 0_{A}\right\rangle \left\vert
0_{B}\right\rangle \left\vert 1_{c}\right\rangle ]
\end{equation}
with the Bell state $\left\vert Bell\right\rangle _{AB}=(\left\vert
1_{A}\right\rangle \left\vert 0_{B}\right\rangle +\left\vert
0_{A}\right\rangle \left\vert 1_{B}\right\rangle )/\sqrt{2}$. Our interest
is in the components of the vacuum state $\left\vert 0_{c}\right\rangle$ of
the WGM, which as shown in Eq. (9) correspond to the Bell state, and we have
to project the WGM on the state $\left\vert 0_{c}\right\rangle $ \cite{exp2}%
. To obtain the Bell state $|Bell\rangle $ with high success rates, we need
to ensure that the coupling strength $g_{0}$ is much larger than the Rabi
frequency of the laser pulse $\Omega_{0}$. So noteworthy features of our
scheme are as follows: (i) In principle, the excited states of NV centers
are negligibly populated due to dark state dominant in the evolution, so the
preparation of the Bell state is immune to spontaneous emission; (ii) The
cavity decay exists only for a short time with the intermediate state $%
\left\vert 1_{c}\right\rangle$ populated, whereas the satisfied condition $%
g_{0}\gg \Omega_{0}$ makes the detrimental influence from the cavity decay
negligible during the operations. We will justify these points by numerics
later.

Note that the method of dark-state evolution can be also applied to quantum
information transfer (QIT) between any pairs of NV centers, where the the
original quantum information is encoded on the NV center $A$ as $\left\vert
\Psi _{0}\right\rangle =c_{0}\left\vert 0_{A}\right\rangle +c_{1}\left\vert
1_{A}\right\rangle$ ($c_{0}$ and $c_{1}$ are arbitrary coefficients), which
can be coherently transferred to the NV center $B$ (initially prepared in $%
\left\vert 0_{B}\right\rangle $ state) via Raman transitions induced by a
pair of time-delayed laser pulses. We can apply such a "counterintuitive"
pulse sequence from $\Omega_{A}(t)/\Omega_{B}(t)\ll 1$ to $\Omega
_{A}(t)/\Omega_{B}(t)\gg 1$, namely, the pulse on NV center $B$ precedes the
pulse on NV center $A$, which guarantees that the adiabatic transfer of the
quantum information could be achieved. Compared with the creation of Bell
state, there is only a slight modification on the step (iii) in above
mentioned procedure, i.e., slowly changing the Rabi frequencies to meet the
condition $\Omega_{A}\gg $ $\Omega_{B}$, rather than $\Omega_{A}=$ $\Omega
_{B}$. In addition, different from in preparing Bell state, the initial
state $\left\vert \Psi _{0}\right\rangle \left\vert 0_{B}\right\rangle
\left\vert 0_{c}\right\rangle $ will drive the system to undergo a different
dark-state evolution involving two dark states $\left\vert D\right\rangle $
and $\left\vert D\right\rangle ^{^{\prime }}$, where $\left\vert
D\right\rangle ^{^{\prime }}=\left\vert 0_{A}\right\rangle \left\vert
0_{B}\right\rangle \left\vert 0_{c}\right\rangle $ is another dark state
regarding the Hamiltonian in Eq. (5). The QIT process could be briefly
expressed as
\begin{eqnarray}
&&(c_{0}\left\vert 0_{A}\right\rangle +c_{1}\left\vert 1_{A}\right\rangle
)\left\vert 0_{B}\right\rangle \left\vert 0_{c}\right\rangle
\underrightarrow{\text{ (ii)\ }}c_{0}\left\vert D\right\rangle ^{^{\prime
}}+c_{1}\left\vert D\right\rangle  \notag \\
&&\underrightarrow{\text{ (iii) }}\left\vert 0_{A}\right\rangle
(c_{0}\left\vert 0_{B}\right\rangle +c_{1}\left\vert 1_{B}\right\rangle
)\left\vert 0_{c}\right\rangle .
\end{eqnarray}

\subsection{Decay Case}

Without losing generality, we consider two NV centers with the identical
parameters and two laser pulses with Gaussian envelops $\Omega
_{A}(t)=\Omega _{m}e^{-(t-\tau _{A})^{2}/\bigtriangleup \tau ^{2}}$ and $%
\Omega _{B}(t)=\Omega _{m}e^{-(t-\tau _{B})^{2}/\bigtriangleup \tau ^{2}}$,
where $\Omega _{m}$ is the maximal value of $\Omega _{(A)B}$ at the central
time $\tau _{j}$ for the pulse $j$ ($j=A,B$), $\bigtriangleup \tau $ is the
laser beam waist. Fig. 5 presents the numerical treatment for the QIT
process: $\left\vert 1_{A}\right\rangle \left\vert 0_{B}\right\rangle
\left\vert 0_{c}\right\rangle \rightarrow \left\vert 0_{A}\right\rangle
\left\vert 1_{B}\right\rangle \left\vert 0_{c}\right\rangle $. We have
compared the population of $\left\vert 0_{A}\right\rangle \left\vert
1_{B}\right\rangle \left\vert 0_{c}\right\rangle $ in an ideal dark-state
evolution described by Eq. (5) with that in the decay case. If no photon
leakage really happens either from the excited state or from the cavity mode
during the gating period, the system is governed by
\begin{eqnarray}
H_{decay} &=&\sum\nolimits_{j=A,B}[g_{j}a^{+}\left\vert 0_{j}\right\rangle
\left\langle e_{j}\right\vert +\Omega _{j}(t)\left\vert e_{j}\right\rangle
\left\langle 1_{j}\right\vert +H.c.]  \notag \\
&&-i\frac{\kappa }{2}a^{+}a-i\frac{\Gamma }{2}\sum\nolimits_{j=A,B}\left%
\vert e_{j}\right\rangle \left\langle e_{j}\right\vert ,
\end{eqnarray}
where $\kappa $ is the cavity decay rate, and $\Gamma$ is the spontaneous
emission rate with respect to the excited state $\left\vert e\right\rangle $
. We set the values of other experimental parameters as $g_{A}=g_{B}=g_{0}=2
\pi \times 1$ GHz, $\Omega _{m}\simeq 2\pi \times 470$ MHz, $\tau _{A}=6.8$
ns, $\tau _{B}=5$ ns, $\bigtriangleup \tau =1.8$ ns, and $\kappa =\Gamma
=g_{0}/10$, which fulfill the adiabatic conditions $\Omega
_{m}\bigtriangleup \tau \gg 1$ and $g_{j}\bigtriangleup \tau \gg 1$ \cite%
{Vit}. In our case, the time delay $(\tau _{A}-\tau _{B})$ between two
pulses $\Omega _{A}$ and $\Omega _{B}$ of the same step is chosen to be
equal to $\bigtriangleup \tau $ to minimize the nonadiabatic losses \cite%
{Sang}. Moreover, the condition $g_{j}\gg \Omega _{m}$ guarantees that the
cavity mode is negligibly populated during the interaction with the pulses,
as demonstrated in Fig. 5.
\begin{figure}[tbph]
\centering
\includegraphics[width=8cm]{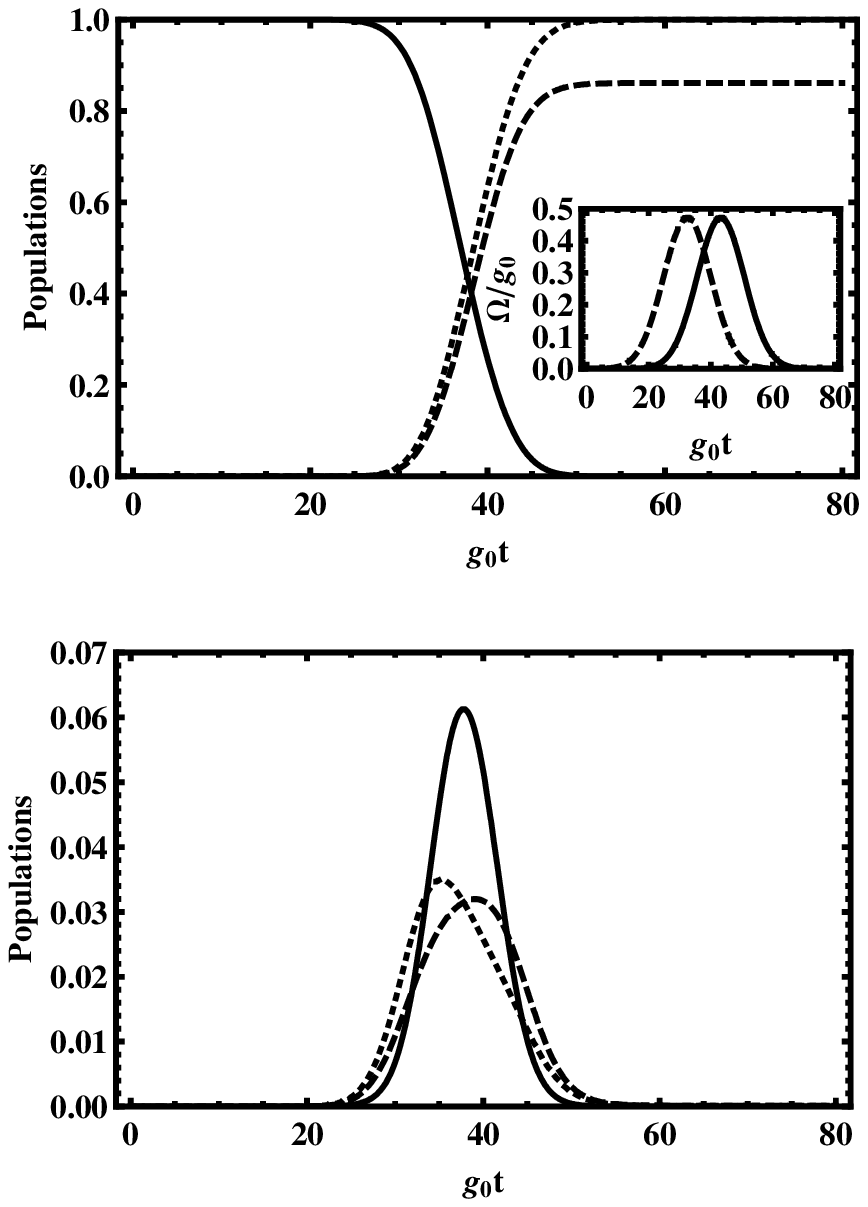}
\caption{Numerical simulation exhibiting the populations versus
$g_{0}t.$ Top: the solid curve denotes the population of $\left\vert
1_{A}\right\rangle \left\vert 0_{B}\right\rangle \left\vert
0_{c}\right\rangle .$ The dotted and dashed curves represent the
population of $\left\vert 0_{A}\right\rangle \left\vert
1_{B}\right\rangle \left\vert 0_{c}\right\rangle $ in the ideal and
decay cases, respectively. The inset shows the Rabi frequencies
$\Omega _{A}$ (solid line) and $\Omega _{B}$ (dashed line). Bottom:
the solid, dotted, and dashed curves represent the populations of
$\left\vert 0_{A}\right\rangle \left\vert 0_{B}\right\rangle
\left\vert 1_{c}\right\rangle $, $\left\vert e_{A}\right\rangle
\left\vert 0_{B}\right\rangle \left\vert 0_{c}\right\rangle $, and
$\left\vert 0_{A}\right\rangle \left\vert e_{B}\right\rangle
\left\vert 0_{c}\right\rangle $, respectively, where the parameters
used are $\Omega
_{m}\bigtriangleup \protect\tau =5$ and $g_{0}\bigtriangleup \protect\tau %
=11.$}
\end{figure}

We emphasize that adiabatic passage method could make the QIT process robust
against the fluctuations of experimental parameters, such as small
variations of the peak Rabi frequencies $\Omega _{m}$, and of the WGM-NV
coupling strength $g_{0}$. Once the adiabatic conditions $\Omega
_{m}\bigtriangleup \tau \gg 1$, $g_{j}\bigtriangleup \tau \gg 1$ and $%
g_{j}\gg \Omega _{m}$ could be well satisfied, our scheme neither requires
accurate manipulation of the intensities of the laser pulses, nor demands
precise control of the WGM-NV interaction. The only thing we need to do is
keeping the phase differences stable between the laser pulses $\Omega _{A}(t)
$ and $\Omega_{B}(t)$, which is easily controllable experimentally. It
should be mentioned that the fidelity of the Bell state $\left\vert
Bell\right\rangle _{AB}$ depends only on the ratio of the WGM-NV coupling
constants $g_{A}$ and $g_{B}$ after the laser pulses $\Omega _{A}(t)$ and $%
\Omega _{B}(t)$ are turned off. If $g_{A}$ and $g_{B}$ are not the same, the
two NV centers will be prepared in the state $(g_{A}\left\vert
1_{A}\right\rangle \left\vert 0_{B}\right\rangle +g_{B}\left\vert
0_{A}\right\rangle \left\vert 1_{B}\right\rangle )/\sqrt{ g_{A}^{2}+g_{B}^{2}%
}$ with the fidelity $F=(g_{A}+g_{B})^{2}/[2(g_{A}^{2}+g_{B}^{2})]$.

\section{Feasibility and challenge}

We examine the feasibility of our scheme and survey the relevant
experimental parameters. As stated above, NV centers are well suited for
cavity QED because they have a long electron spin relaxation time and their
electronic states can be initialized, manipulated, and measured through
highly stable optical and microwave excitations at room temperature.
Additionally, WGM microcavities exhibit exciting characteristics such as
extremely high-quality factor $Q$, small mode volume $V_{m}$, and excellent
scalability, which makes it possible to achieve high concentration of the
optical field and relatively long photon confinement times. Experimentally,
single N-V centers strongly coupling to a WGM have been demonstrated in
different kinds of microcavities \cite{park,sch,bar}, which are great
advances in WGM-based cavity QED research.

In the composite nanocrystal-microsphere system, we focus our attention on
the dipole transition $\left\vert 0\right\rangle \leftrightarrow \left\vert
e\right\rangle $ with a ZPL at $\lambda =637$ $nm$ of the NV centers in
diamond nanocrystal. In above-mentioned WGM-NV experiments, the NV centers
actually interact with the evanescent field of the WGM. The evanescent field
is of importance because it offers the effective way to energy exchange
between the WGM and the external NV center. Thus, we assume that single NV
centers are located near the microcavity surface in order for the WGM-NV
coupling $g_{\max }$ to reach $2$ $\pi\times 1$ GHz \cite{WG} with the
parameters in the Sec. 2.2. In the fused-silica microsphere cavity, the
small radius of 10 $\mu$m could lead to a vacuum electric field of $150$
V/cm at the sphere surface and to the $Q$ factor exceeding 10$^{9}$, which
imply the WGM decay rate to be $\kappa =\omega _{e0}/Q=2\pi \times 0.47$
MHz. Experimental studies have demonstrated $Q$ factors approaching $10^{10}$
in a silica WGM microsphere \cite{Ar,ver}, with values exceeding 10$^{8}$
readily achievable over a broad range of cavity diameters and wavelengths.
The strong coupling strongly depends on the critical photon number $%
n_{0}=\Gamma _{0}^{2}/(2g_{\max }^{2})$ and the critical atom number $%
N_{0}=2\Gamma _{0}\kappa /(g_{\max }^{2})$, which gives the number of
photons required to saturate a NV-qubit, and the number of NV-qubit required
to have an appreciable effect on the WGM cavity transmission, respectively
\cite{WG}. Based on these parameters, one can find that the strong coupling
conditions $g_{\max }\gg$ $\kappa, \Gamma_{0}$ and $(n_{0}, N_{0})\ll 1$
could be well satisfied in our scheme.

Nevertheless, in realistic experiments, WGMs are dominantly confined inside
of the microcavity body, and only the remaining energy of the WGM can be
resident in the exterior evanescent field. Thus the coherent coupling
strength cannot reach its maximum. This is the reason that the maximal
WGM-NV coupling in current experiments only reaches $2\pi \times 300$ MHz
\cite{bar}. There is an increasing interest in the development of
alternative microcavity systems. We note that a recent experiment \cite{XY1}
has demonstrated an enhanced coherent interaction between the WGM and
quantum dots using a kind of plasmonic WGM highly localized on the exterior
surface of a metal-coated microtoroidal. Another alternative soultion is
using the silica microtoroidal coated with a high-refractive-index (HRI)
nanolayer \cite{XY2}. The key idea is that this HRI nanolayer can compress
the radial WGM field and move the WGM field in the coated microtoroidal to
the outside of the silica surface.

Note that the preparation of entangled states and implementation of QIT via
dark states are similar in spirit to the other atom \cite{Parkins} and
atom-like systems \cite{Dot}. However, we consider here a different system,
and our proposal has several merits as follows. Firstly, the NV center in
diamond is the only currently known viable solid-state qubit at
room-temperature, and the center's highly localized bound states are well
isolated from sources of decoherence. So the ground state can exhibit
extremely long spin coherence time of up to \textit{millisecond}, which is
close to the regime needed for quantum error correction \cite{PNAS}.
Secondly, the proposed QIT protocol has the potential of scalability because
our QIT protocol does not require identical WGM-NV coupling strength, which
implies that neither identical qubits nor exact placement of NV centers in
cavities is needed. Thirdly, as the NV center nanocrystals are required to
be attached along the equator of the microsphere with spacing bigger than
the laser wavelength, individual addressing is not an obstacle in our scheme.

The two schemes above require different conditions in implementation. For
example, in the first scheme the WGMs are detuned from the transitions in
the NV centers, while the second scheme requires resonant coupling between
the WGMs and the NV centers. If we employ the same WGM cavities to
accomplish the schemes, the detuned and the resonant couplings regarding the
WGM radiation could be achieved by changing the temperature of the system
\cite{park}.

In current experiments, the electron spin relaxation time $T_{1}$ of diamond
NV centers ranges from 6 ms at room temperature \cite{Neu1} to sec at low
temperature. In addition, the dephasing time $T_{2}=$350 $\mu$s induced by
the nearby nuclear-spin fluctuation has been reported \cite{ini1}. A latest
experimental progress \cite{latest} with isotopically pure diamond sample
has demonstrated a longer T$_{2}$, i.e., T$_{2}=$2 ms. In our scheme, the
operation times for preparing four-qubit W state, the two-qubit Bell state,
and for accomplishing QIT are about 31.3 ns, 6.786 ns, and 8.0 ns,
respectively. Hence even at room temperature, up to $10^{3}\sim 10^{4}$ gate
operations are feasible under the present experimental conditions.

For more technical aspects, to make sure the NV centers to be strongly and
nearly equally coupled to the WGM cavity, we have to simultaneously attach
separate NV centers around the equator of a single fused-silica microsphere
resonator. The experimental difficulty lies in how to fix the nanocrystals
appropriately with respect to the WGM. So far the strong coupling between a
single NV center and the WGM of a silica microsphere \cite{park} and a
gallium-phosphide microdisk \cite{bar}, and between two NV centers and a
polystyrene microsphere \cite{sch} have been experimentally achieved. So we
wish this would be soon extended to more NV centers. Further consideration
would involve nuclear spins in the NV centers, which are more suitable to
store quantum information due to longer decoherence time. In this sense, we
may also consider quantum computation with NV centers combining nuclear
spins with electron spins, i.e., encoding the qubits in the nuclear spins
and employing the electron spins as ancillas. The hyperfine interaction
helps to transfer the generated entanglement of the electron spins to the
corresponding nuclear spins of NV centers \cite{register}.

\section{Conclusion and Acknowledgements}

In conclusion, we have proposed two schemes to prepare W state and Bell
state with separate NV centers in the diamond nanocrystal-microsphere
system, respectively. In both of these schemes, the cavity decay and the
spontaneous emission from the excited states could be effectively
suppressed. Particularly, in the latter scheme with the adiabatic passage,
the QIT is robust against the experimental parameter fluctuation.

Our ideas could be applied to other cavity systems, besides the WGM-type
cavities. The number of the entangled NV centers depends on the ratio of the
size of the employed microcavity to the nanocrystal size. For more NV
centers to be entangled, detection of emitted photons by parity projection
would be necessary \cite{Review}. In this sense, we argue that our present
work has demonstrated a building block for a large-scaled NV center system,
which would be feasible in the near future.

WLY thanks Zhang-qi Yin for enlightening discussions. This work is supported
by National Natural Science Foundation of China under No. 10974225, by
Chinese Academy of Sciences, and by 973 project.


\begin{thebibliography}{99}
\bibitem{exp1} The spontaneous decay rates from $\left\vert e\right\rangle $
to $\left\vert 1\right\rangle $ and from $\left\vert e\right\rangle $ to $%
\left\vert 0\right\rangle$ are assumed to be equal.
\bibitem{exp2} The WGM state could be detected by dint of an auxiliary NV
center initially in the state $\left\vert 0\right\rangle $. The transition $%
\left\vert 0\right\rangle \iff \left\vert e\right\rangle $ is resonant with
the WGM for an interaction time $t=\pi /2g$ with $g$ the coupling strength
due to the WGM. If the auxiliary NV center is detected in the state $%
\left\vert 0\right\rangle$, the state of WGM was initially in the $%
\left\vert 0_{c}\right\rangle$ state.
\end{thebibliography}
\end{document}